\documentclass[9pt,twocolumn,twoside]{opticajnl}
\journal{opticajournal} 

\setboolean{shortarticle}{true}

\usepackage{lineno}

\usepackage{graphicx}
\usepackage{dcolumn}
\usepackage{bm}

\usepackage[utf8]{inputenc}
\usepackage[T1]{fontenc}
\usepackage{mathptmx}
\usepackage{etoolbox}
\usepackage{subcaption}

\title{Wafer-scale fabrication of evacuated alkali vapor cells}

\author[1,2, a)]{Yang Li}
\author[1,2,3, a)]{Donggyu B. Sohn}
\author[1]{Matthew Hummon}
\author[1]{Susan Schima}
\author[1]{John Kitching}

\affil[1]{Time and Frequency Division, National Institute of Standards and Technology, Boulder, CO USA}
\affil[2]{Department of Physics, University of Colorado, Boulder, CO USA}
\affil[3]{Now at PsiQuantum, Palo Alto, California 94304}
\affil[a)]{These authors contributed equally to this work.}

\begin{abstract}
We describe a process for fabricating a wafer-scale array of alkali metal vapor cells with low residual gas pressure. We show that by etching long, thin channels between the cells on the Si wafer surface, the residual gas pressure in the evacuated vapor cell can be reduced to below 0.5 kPa (4 Torr) with a yield above 50 \%.  The low residual gas pressure in these mass-producible alkali vapor cells can enable a new generation of low-cost chip-scale atomic devices such as vapor cell optical clocks, wavelength references, and Rydberg sensors. 
\end{abstract}

\setboolean{displaycopyright}{false} 

\begin{document}

\maketitle
Vapor cells containing alkali atoms have been a cornerstone of atomic physics experiments for many decades, enabling the demonstration of phenomena from optical pumping \cite{happer1972} to laser cooling \cite{monroe1990very}. They also are the heart of many atomic clocks \cite{Arditi1960, Carpenter1960} and quantum sensors such as magnetometers \cite{Dehmelt1957, Bell1957}  and electric field sensors based on Rydberg atoms \cite{Sedlacek2012}. For sensors based on spin transitions in atoms, such as microwave clocks and magnetometers, buffer gases are added to the cell along with the alkali atoms in order to reduce wall-induced collisional decoherence of the atoms and improve the transition quality factors. However, such gases typically create strong perturbations to the atoms' electronic transitions, making them largely unsuitable for instruments such as optical frequency references \cite{hummon2018photonic} and Rydberg-based sensors \cite{Sedlacek2012}. For these latter instruments, cells containing only alkali (or alkaline-earth) atoms with no residual gases are desirable.

Over the last 20 years, fabrication processes have been developed to create compact vapor cells using etched silicon and glass \cite{liew2004microfabricated, kitching2018chip}. These cells enable commercially available \cite{lutwak2003chip,lutwak2011sa} instruments such as chip-scale atomic clocks \cite{knappe2004microfabricated} and magnetometers \cite{schwindt2004chip} and are a key enabling factor for small size and low power consumption. Methods to reduce fabrication cost and increase yield and uniformity from cell to cell are therefore of high importance. Prior work \cite{bopp2020wafer} demonstrated a new method for fabricating large wafers of vapor cells using a single process sequence. In this method, chemical precursors for the alkali metal were deposited into an array of holes on a silicon wafer, and the resulting alkali metal was then transported into a second set of holes through lateral translation between the silicon wafer and the glass lid wafer before bonding. This process successfully created cells containing buffer gases, but considerable residual gases, as much as several kPa (several tens of Torr), were present in cells intended to be evacuated.

In this letter, we describe a wafer level process for the fabrication of evacuated alkali metal vapor cells with considerably improved yield and uniformity over previous work. This is accomplished by the introduction of thin channels etched into the wafer surface between the cells to allow gases to escape during the fabrication process. Using optical spectroscopy to characterize the cells, we find that more than 50 \% of the cells on the wafer have residual gas pressures below  0.5 kPa (4 Torr), consistent with zero within our measurement uncertainty.

\begin{figure*}[!htb]
  \centering
    \begin{subfigure}[b]{0.4\textwidth}
     \includegraphics[width=\textwidth]{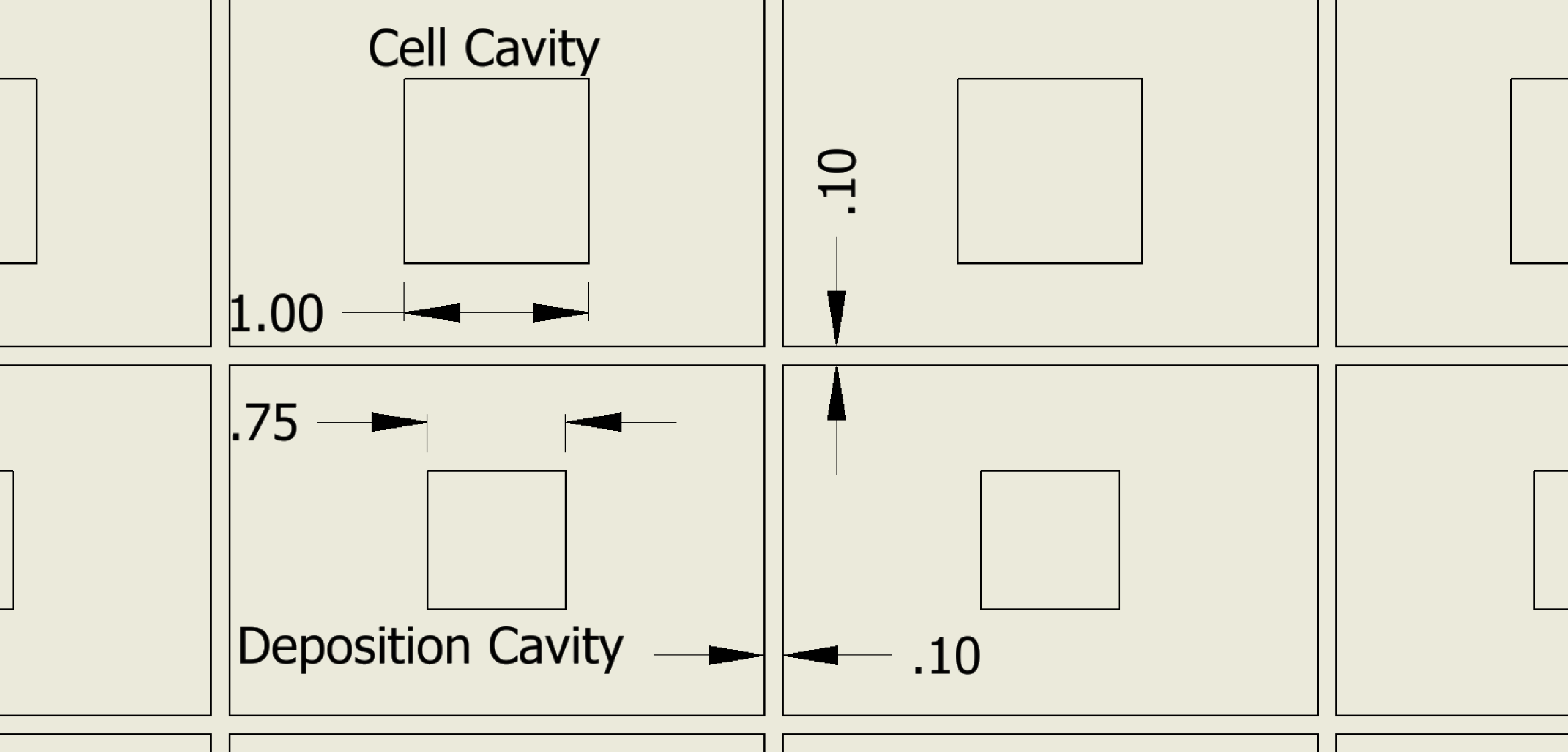}
     \subcaption{\label{fig:cell_dimension}}
      \end{subfigure}
        \begin{subfigure}[b]{0.4\textwidth}
     \includegraphics[width=\textwidth]{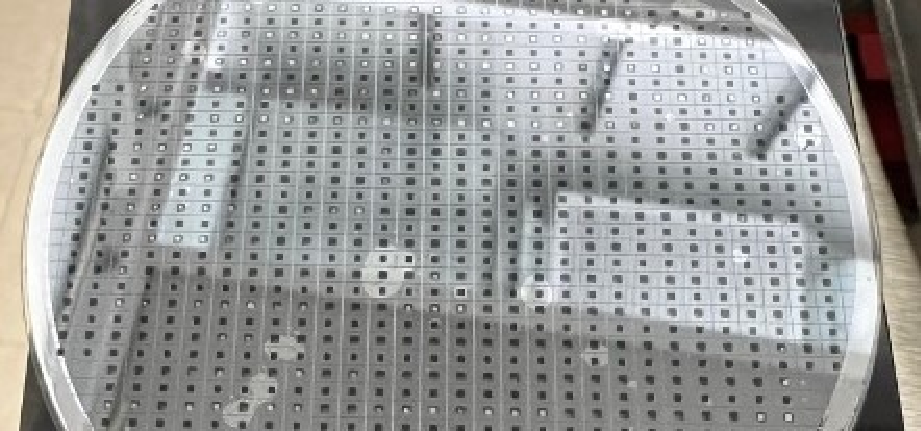}
      \subcaption{\label{fig:cell_picture}}
      \end{subfigure}
     \caption{ (a) Schematic of cell cavities, deposition cavities and channels running between them. Dimensions are in mm. (b) Photograph of the 4 inch wafer after fabrication showing the channels, which run to the edge of the wafer. The shiny spots that can be seen in some cells are metallic Rb.  }
    \label{fig:cell_wChannel}
\end{figure*}

The main fabrication steps in the process have been summarized previously in Bopp et al.\cite{bopp2020wafer}. An array of hole pairs is etched through a Si wafer 1 mm thick and four inches in diameter using deep reactive ion etching (DRIE).  The bigger "cell" cavities form the vapor cells that will contain the alkali metal after the fabrication is complete. The smaller "deposition" cavities contain the chemical precursors for the alkali metal and are ultimately discarded. In this letter, 
 we etched channels 100 $\mu$m deep and 100 $\mu$m wide between the cavities on the front side of the wafer as shown in Fig. \ref{fig:cell_dimension}. These channels run all the way to the edge of the wafer, pass within 1 mm of the cell sides and are intended to allow gases generated during the final anodic bonding step to escape from the cell cavities.

After etching, the unpatterned back side of the Si wafer is anodically bonded to a borosilicate glass wafer in a commercial wafer bonder to form a wafer preform. Precursor materials BaN$_6$ and RbCl are dissolved in water and dispensed into the deposition cavities on the wafer preform with an automated fluid dispensing apparatus. Only the central 15x24 region of the full 21x30 array was filled with precursor materials due to limitations in the automated dispensing. The preform is then loaded into the lower platen of the bonder and a glass capping wafer is loaded into the upper platen. The bonding chamber is then evacuated to below 10$^{-5}$ mbar. The lower wafer preform is placed in contact with the glass capping wafer and heated to 200 ${}^\circ$ C. At this temperature, the precursor materials react (BaN$_6$ $\rightarrow$ Ba + 3N$_2$; Ba + RbCl $\rightarrow$ Rb + BaCl) and Rb metal vapor is released into the deposition cavities.

\begin{figure}[!ht]
    \centering
    \includegraphics[width=0.4\textwidth]{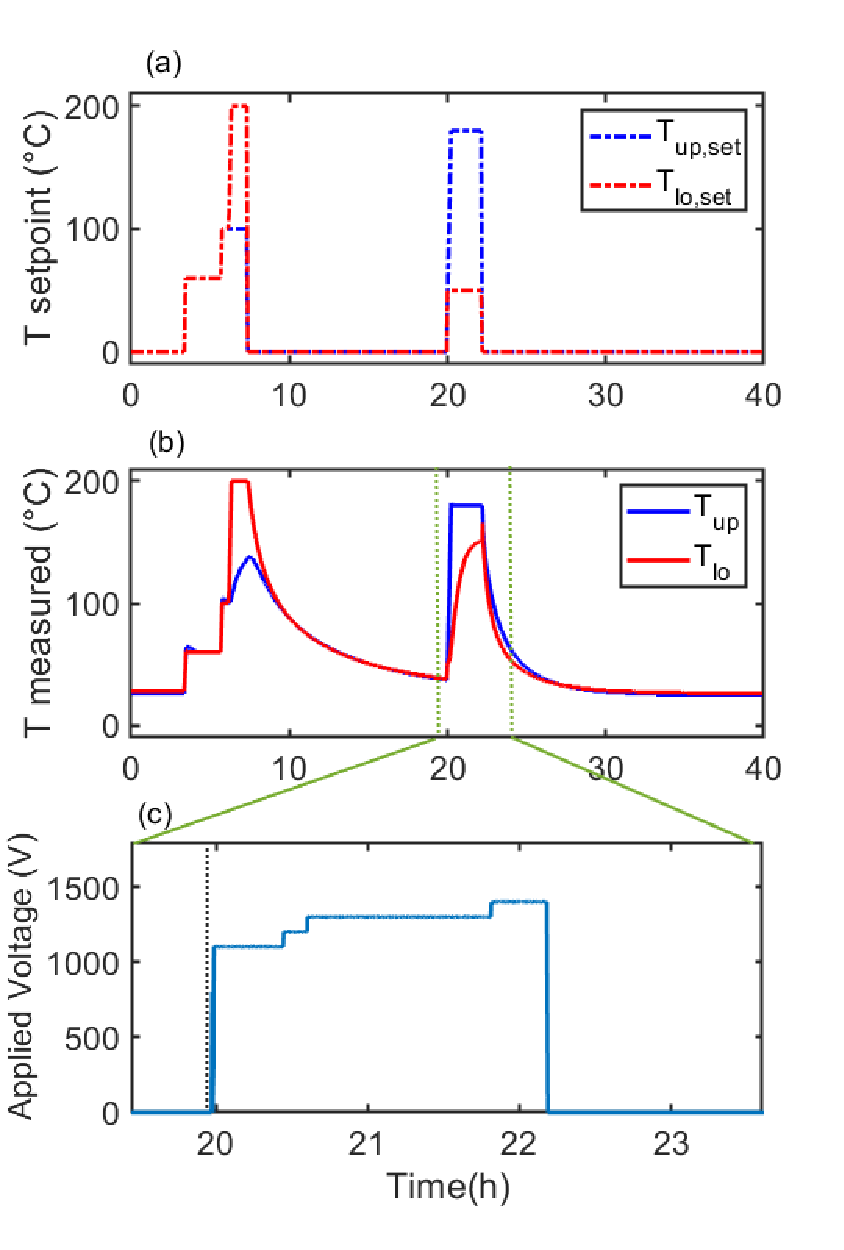}
    \caption{(a) Temperature set points for the upper (blue) and lower (red) platens throughout the fabrication process. (b) Measured temperatures for the upper (blue) and lower (red) platens. The first period of elevated temperature (between 3 and 8 hours) is the deposition step when the precursor chemicals are reacted and alkali metal is created. The second period of elevated temperature (between 19 and 23 hours) is when the final bonding occurs.  (c) Applied bonding voltage during the bonding step. The wafers are moved with respect to each other at around 20 h, indicated by the black dashed line. The stepped ramping of temperature and voltage were made to ensure there were no problems with arcing and are not essential to the process. }
    \label{fig:aml_log}
\end{figure}
The lower platen is maintained at 200${}^\circ$ C for about 1 hour, as shown in Fig. \ref{fig:aml_log}a, so that the Rb condenses on the inside surface of the top glass wafer. The upper platen is not independently heated but increases in temperature over time due to the contact with the heated lower platen, as shown in Fig. \ref{fig:aml_log}b. After the reaction is finished and enough Rb is collected on the top glass, the heater current is shut off and the wafers are allowed to cool for about 12 hrs until both wafers have reached room temperature. The wafers are then separated under vacuum and Rb on the upper glass is transferred to cell cavities through lateral motion of the two wafers. Finally, the upper glass wafer is heated to 180 ${}^\circ$ C, while applying a voltage of around 1300 V between the upper glass wafer and the silicon layer on the lower wafer stack (see Fig. \ref{fig:aml_log}c.). This  final anodic bonding step seals Rb in the cell cavities. 

\begin{figure}[!hb]
    \centering
    \includegraphics[width=0.4\textwidth]{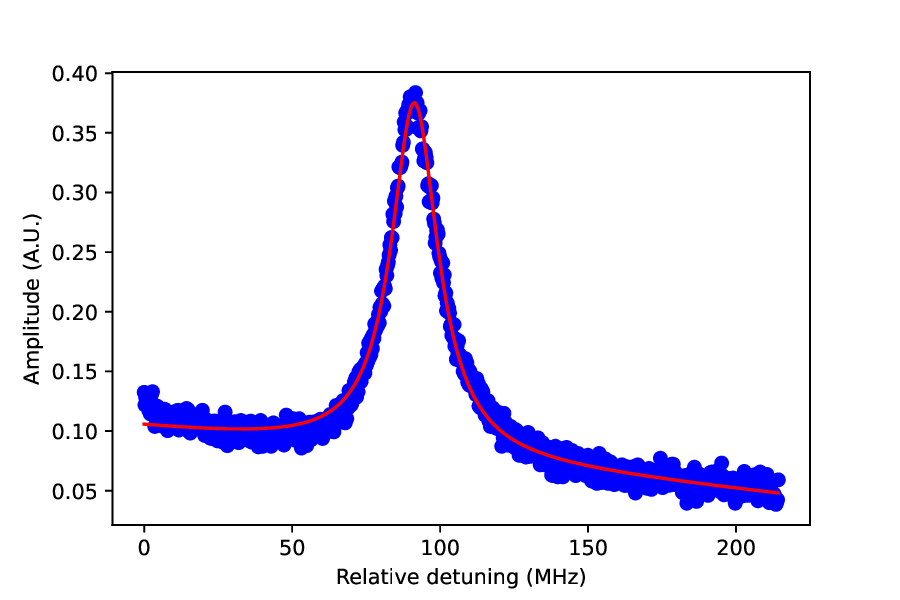}
    \caption{  Saturated peak of the $^{87}$Rb D2 line after removing the Doppler background by the lock-in amplifier. Pump power is 467 $\mu$W, probe power is 12 $\mu$W. The red line shows the best fit to the spectrum  (see text). }
    \label{fig:sas_peak}
\end{figure}

During the final anodic bonding, residual gases released by the continuing BaN$_6$ decomposition and gases generated by anodic bonding can be trapped in the vapor cell if not pumped out efficiently \cite{bopp2020wafer}.  The channels between the cavities provide a higher-conductance path from the cell interiors to the chamber vacuum than exists due to the thin gap between the wafer surfaces when no channels are present.  The wafer after fabrication is shown in Fig.\ref{fig:cell_picture}.

\begin{figure*}[!htb]
  \centering
    \begin{subfigure}[b]{0.4\textwidth}
     \includegraphics[width=\textwidth]{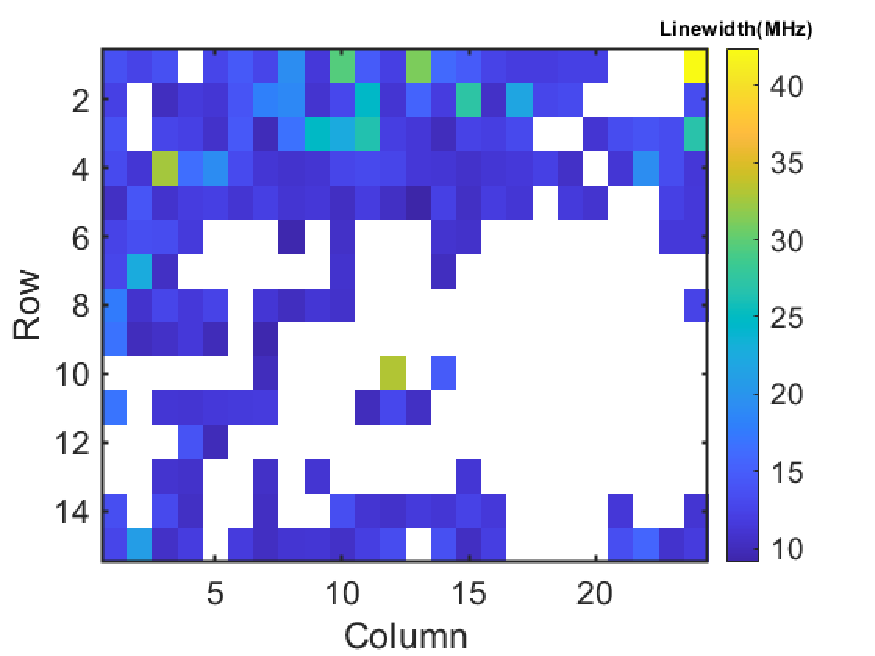}
     \subcaption{\label{fig:Lzpw_colormap}}
      \end{subfigure}
      \begin{subfigure}[b]{0.4\textwidth}
     \includegraphics[width=\textwidth]{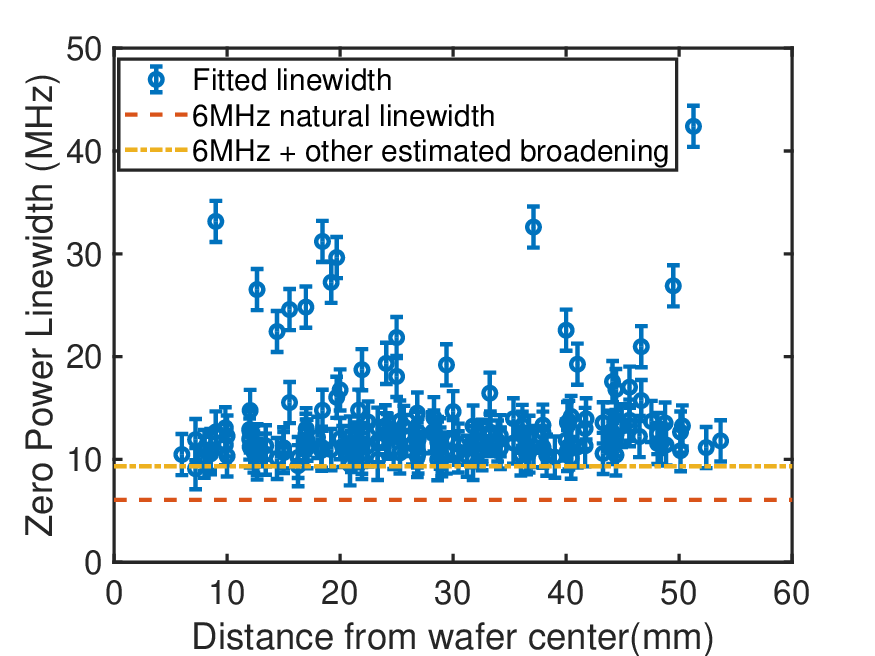}
      \subcaption{\label{fig:Lzpw_uncert}}
      \end{subfigure}
     \caption{ (a) Color map of zero power linewidth. (b) Linewidth as a function of distance from wafer center. The 6 MHz natural linewidth and other estimated broadening from magnetic field, laser linewidth, and transit time broadening are shown as dashed lines.}
    \label{fig:Lzpw}
\end{figure*}

Saturation absorption spectroscopy (SAS) was performed for all the cells on the wafer that contained precursor material at $\approx$ 70 ${}^\circ$C to infer the residual gas pressure. The pump and probe beam from a distributed Bragg reflection (DBR) laser tuned to the $^{87}$Rb D2 line at 780 nm were aligned to overlap within each cell. Lock-in detection of the probe beam while chopping the pump beam removed the Doppler background of the absorption spectrum and the width of the well-isolated F$_g$ = 2 $\rightarrow$   F$_e$ = 3 peak was measured to assess the presence of residual gases in each cell. An example spectrum from one of the cells is shown in Fig.\ref{fig:sas_peak}. The frequency axis is calibrated from the known separation between the saturated peaks (not shown in Fig.\ref{fig:sas_peak}). The heated wafer was mounted on a computer-controlled moving stage which moved the wafer along the x and y axes so each cell could be measured independently. 

The measured lineshapes were fitted by summation of a Lorentzian function, a linear term and a constant offset. The amplitude, linewidth and line center of Lorentzian function are free parameters to be adjusted by the fit, as well as a linear slope and an overall offset.  The linewidths for each cell were measured at five different pump powers, which allowed a determination of the zero-power linewidth  $\Gamma_0$ and standard error according to $\Gamma = \Gamma_0 (1 + I/I_S)^{1/2}$, where $\Gamma$ is the linewidth from the Lorentzian fit, $I$ is the pump intensity, and $I_S$ is the saturation intensity \cite{allen1987optical}.

The zero-power linewidths extracted from the central 15x24 cells on the wafer are shown in Fig.\ref{fig:Lzpw_colormap}. The white regions indicate cells with no observable absorption, and therefore containing no Rb. There is a cluster of cells on one side of the wafer that have larger linewidths, suggesting that in this region, residual gases were insufficiently pumped through the channels. The zero-power linewidth as a function of distance from the center of the wafer is shown in Fig.\ref{fig:Lzpw_uncert}; no obvious trend is observed, indicating that the channels enable efficient pumping of the cells independent of their distance to the wafer edge. Channels reduce the pumping path length of residual gases from up to 50 mm for cells in the center of the wafer to below 1 mm.  The fitting error bar is 2 MHz in Fig.\ref{fig:Lzpw_uncert}. 

The zero-power linewidth includes contributions from natural linewidth of $^{87}$Rb (6.07 MHz) shown as the dashed line in Fig. \ref{fig:Lzpw_uncert}, pressure broadening caused by residual gas in the cell, magnetic field broadening, drift of the laser during the scan, and the laser linewidth. From rough measurements of the magnetic field near the cells, we estimate the broadening due to magnetic field to be 2 MHz, and the linewidth of the DBR laser is about 1 MHz. The transit time broadening is estimated to be 270 kHz for a 1 mm laser beam.  A conservative upper limit on the collisional broadening caused by the residual gas pressure can therefore be inferred  by subtracting only the 6 MHz natural linewidth from the total broadening.

\begin{figure}[!ht]
    \centering
    \includegraphics[width=0.49\textwidth]{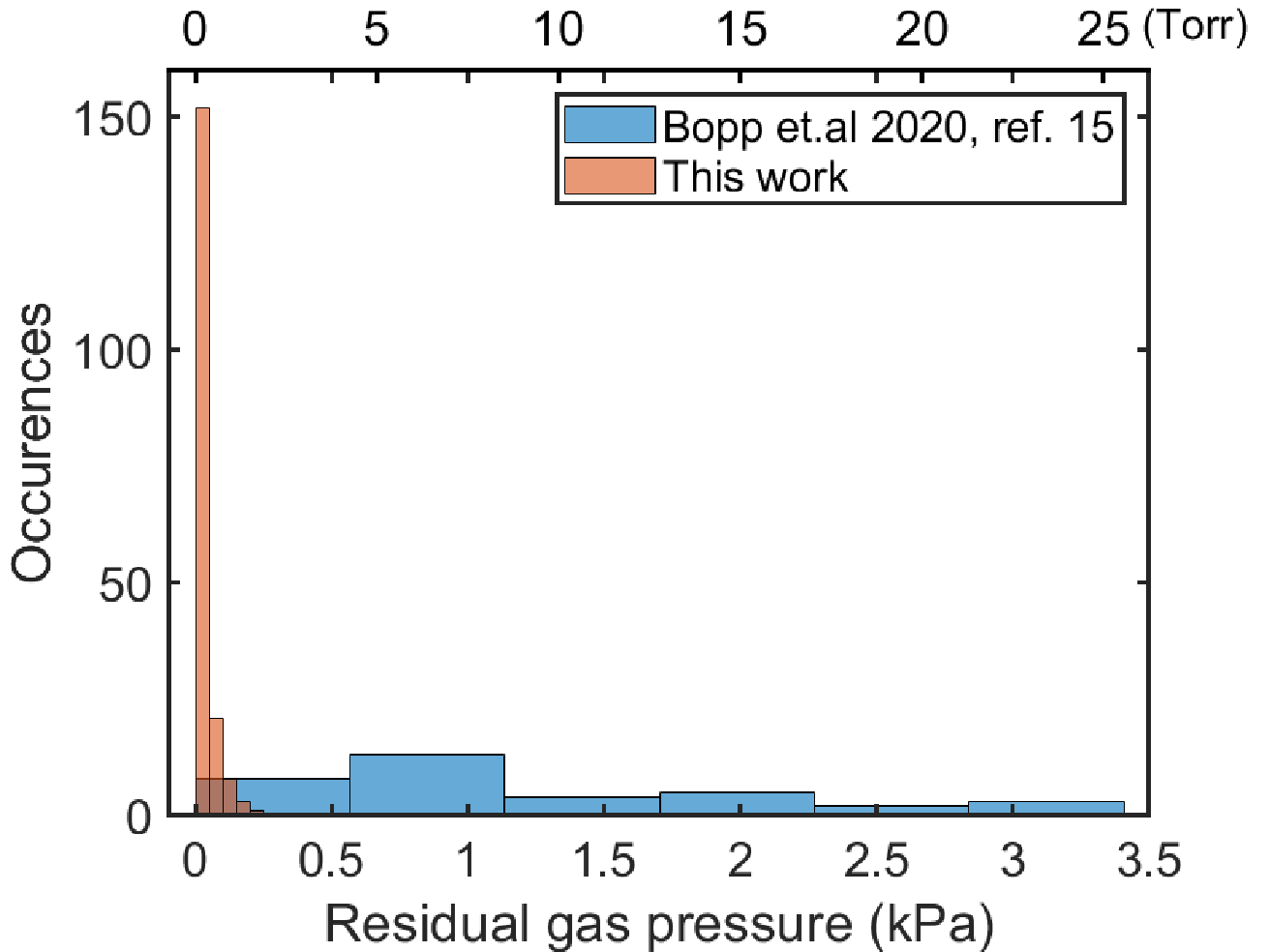}
    \caption{ Histogram of residual gas pressure upper bound for the wafer with channels, compared to the wafer without channels \cite{bopp2020wafer}.  The presence of the channels results in a residual gas pressure smaller than  0.5 kPa (4 Torr) in the vapor cells.}
    \label{fig:histograph}
\end{figure}

We use a collisional broadening  coefficient of 18.1  GHz/amg\cite{romalis1997pressure} to convert the residual broadening into a residual gas pressure. Fig. \ref{fig:histograph} shows the residual gas pressure histogram for the channel wafer compared with a previous wafer with no channels \cite{bopp2020wafer}.  The presence of the channels reduces the residual gas pressure to below 0.5 kPa (4 Torr), which is about ten times lower than the residual gas pressure in the previous method \cite{bopp2020wafer} without channels. Because of the presence of additional broadening mechanisms (due to magnetic fields, laser linewidth, etc.), the estimated residual gas pressure in the majority of the cells is consistent with zero. We also found that since the residual gases are quickly pumped away during the precursor decomposition step, Rb reaches the upper glass wafer more quickly, which in turn helps to improve the yield. 

In conclusion, we have demonstrated a process for fabricating evacuated alkali vapor cells in large wafers by introducing 100 $\mu$m deep channels across the Si wafer surface. A residual gas pressure smaller than 0.5 kPa (4 Torr) is achieved, consistent with zero, with a yield of 51\% referenced to the number of cells initially filled with precursor materials. The residual pressure in such cells is already sufficient for optical references based on Doppler-free transitions in alkali atoms. More sensitive spectroscopy is needed to assess whether the residual gas pressures are sufficiently low for high-performance Rb two-photon optical references \cite{RN9303} or Rydberg-atom-based quantum sensors \cite{Sedlacek2012}.

\begin{backmatter}
\bmsection{Funding}  This work was supported by funding from the National Institute of Standards and Technology.

\bmsection{Acknowledgments} The authors thank Marlou Slot and Aly Artusio-Glimpse for helpful comments. 

\bmsection{Disclosures} The authors declare no conflicts of interest.

\bmsection{Data Availability Statement} Data underlying the results presented in this paper are
not publicly available at this time but may be obtained from the authors upon
reasonable request.

\end{backmatter}

\bibliography{Optica-journal-template}



\end{document}